\documentclass[10pt,amsmath,amssymb,epsfig,aps,pra,twocolumn]{revtex4-1}
\usepackage{ifpdf}
\usepackage{graphicx}
\usepackage{amssymb,bm}
\usepackage{amsmath}
\usepackage{color}
\usepackage{xcolor}
\usepackage{multirow}
\usepackage{afterpage}
\usepackage{hyperref}
\usepackage{ulem}
\usepackage{cancel}
\begin{document}
\title[]{Modulational instability of inter-spin-orbit coupled Bose-Einstein condensates in deep optical lattice}

\author{R. Sasireka$^1$, S. Sabari$^2$,  A. Uthayakumar$^1$, and Lauro Tomio$^{2}$}
\address{$^{1}$Department of Physics, Presidency College (Autonomous), 
Chennai - 600005, India.\\
$^{2}$Instituto de F\'\i sica Te\'{o}rica, UNESP -- Universidade Estadual Paulista,
01140-070 S\~{a}o Paulo, Brazil. }
\date{\today}
\begin{abstract}
We present a comprehensive study of modulational instability (MI) in a binary 
Bose-Einstein condensate with spin-orbit coupling, confined to a deep optical lattice.
The system is modeled by a set of discrete Gross-Pitaevskii equations. 
Using linear stability analysis, we derive the explicit MI conditions for the system,
elucidating the critical and distinct roles played by spin-orbit coupling, 
inter-species nonlinearity, and intra-species nonlinearity. Our analysis, 
conducted for both unstaggered and staggered fundamental modes, 
reveals markedly different instability landscapes for these two configurations.
The analytical predictions are confirmed by extensive numerical simulations
of the full nonlinear dynamics, which vividly illustrate the spatiotemporal
evolution of wave amplitudes, phase coherence, and energy localization during
the instability process. The numerical results, obtained via a fourth-order 
Runge-Kutta method, show excellent agreement with the linear stability theory
and provide a complete picture of the MI-induced pattern formation.
\end{abstract}


\date{\today}
\maketitle

\section{Introduction}
\label{sec1} 
Spin-orbit (SO) and Rabi coupling studies in Bose-Einstein condensates (BECs) have 
received strong encouragement after the experimental realization of  {\it synthetic}
spin-orbit coupling (SOC) reported in Ref.~\cite{Lin2011}. Observed in this work, 
the relevance of studying SOC relies on understanding several important effects in 
quantum mechanics, such as the  previously observed spin-Hall effect in 
superconductors and quantum wells~\cite{2004Kato,2007Konig}, which can lead to the
development of spintronic devices. In particular,  ultracold atoms provide an ideal 
platform to study SOC, considering the possible precise experimental control on the 
atom-atom interactions. In Ref.~\cite{Lin2011}, the SOC is {\it engineered} with equal 
Rashba~\cite{1959Rashba,1984Bychkov} and Dresselhaus~\cite{1955Dresselhaus} 
strengths in a neutral atomic BEC by dressing two atomic spin states with a pair 
of lasers, within an approach shown to be equally applicable for bosons 
and fermions. The Rashba SOC is typically obtained by polarized counter-propagating lasers; whereas
the Dresselhaus SOC is obtained by adjusting the relative phases and orientations of the lasers.
The procedure to create SOC is analogous to the one when considering the spin states of 
an electron, by selecting two internal states from the atom, which are labeled as pseudo-spin-up 
and pseudo-spin-down. In~\cite{Lin2011}, by considering the rubidium $^{87}$Rb atom, 
the respective selected states were $|+\rangle\equiv |{\rm F=1, m_F=0}\rangle$ (pseud-spin-up)
and  $|-\rangle\equiv |{\rm F=1, m_F=-1}\rangle$ (pseudo-spin-down).
One of the most attractive and convenient aspects of studying SOC in BECs is 
on the possibilities of tuning the nonlinear interactions, which can be adjusted from repulsive 
to attractive by using Feshbach resonance techniques~\cite{1998Inouye,1999Timmermans}. So,  
plenty of platforms emerge to explore the interplay between nonlinear interactions with the 
linear couplings and possible applied fields to the condensed 
atoms~\cite{2005Ruseckas,2012Edmonds,2012Zhang,2018Abdullaev-Brtka}.
Among the linear couplings, the synthetic SOC can be engineered under different experimental 
conditions for the couplings between two different hyperfine states of an atom, as we aim 
to explore in this work. An interesting approach being pursued concerns SOC in BECs confined 
in periodic optical lattice (OL) potentials. The main relevant aspects related to 
BEC confined in OL, with an overview of the pioneering works, can be found in 
Ref.~\cite{Morsch2006}.
Following that, great progress has been verified, considering various proposed schemes 
to explore SOC with OL potentials, as exemplified in Refs.~\cite{Struck2014,Wall2016}. 

For realizing SO coupling in OLs, the most efficient method in use is applying light-assisted 
tunneling via one-photon or two-photon Raman processes. The two-photon Raman transition couples 
either the hyperfine levels or sublattice sites to the momentum via light-assisted tunneling in 
optical super-lattices~\cite{Jaksch2003,Huang2016,Li2017}, optical Raman
lattices~\cite{Liu2014,Wu2016}, and moving lattices~\cite{Hamner2015}. Alternatively, 
a shaken OL in the context of Floquet engineering is an alternative route for introducing SO 
coupling into OLs~\cite{Struck2014}, but a synthetic magnetic flux has only been implemented
successfully in OLs~\cite{Miyake2013,Jotzu2014}.
Also, the combinations of OL potential with the interaction of SO coupling were shown to 
possess interesting phenomena like flattening of Bloch potential~\cite{Zhang2013}, atomic
Zitterbewegung~\cite{Larson2010}, and new topological phases~\cite{Stanescu2009}. Generally, 
BECs become fragmented on each lattice when trapped by deep OL potential. Such a system 
has been effectively described in the tight-binding approximation by the discrete version of 
the corresponding Gross-Pitaevskii (GP) equation~\cite{Trombettoni2001,Kevrekidis2009,Gligoric2014},
with composite solitons and localized modes being carried out on Rashba-type 
SOC-BEC trapped in deep OL~\cite{Sakaguchi2014b,2015Belicev}. More recently, 
some of us~\cite{Sabari2021} have studied modulational instability in coupled BECs confined 
in deep OL by considering intra-SO coupling. 

Modulational instability (MI) is a generic phenomenon leading to large-amplitude periodic waves, 
which occurs in dynamic systems, like fluids, nonlinear optics, and plasmas. It results from the 
interplay between nonlinear dynamics and dispersion (or diffraction in the spatial domain), with 
the fragmentation of carrier waves into trains of localized waves~\cite{1967Benjamin,1970Hasegawa}, 
corresponding to the growth of weakly modulated continuous waves in a nonlinear medium. Considering 
experimental setups, it was also reported recently investigations in optics and
hydrodynamics~\cite{2021Vanderhaegen}, demonstrating that MI can be a more complex phenomenon than the 
one predicted by the conventional linear stability analysis~\cite{2020Conforti}, which goes 
beyond the limited predicted frequency range.
Experimental realizations of MI in BECs were reported 
in Refs.~\cite{2017Nguyen,2017Everitt}, considering
cigar-shaped trapped condensates, which have indicated 
the relevance of MI in cold-atom physics.
By following some other investigations reported in 
Refs.~\cite{2002Khawaja,2002Strecker,2004Carr}, 
within the nonlinear Schr\"odinger (NLS) formalism, as
the GP equation, plenty of other studies considering MI
analyses have been performed in the last two decades. 
Among them, we can mention some previous contributions
performed by some of us on MI, such as the variational 
analysis considering cubic-quintic NLS
formalism~\cite{Sabari2020}, following 
Ref.~\cite{2007Ndzana}. Some other related studies were, 
for example, by assuming
scalar~\cite{2013Sabari,2015Sabari,2014Wamba,Sabari2022} and
vector~\cite{Goldstein,Kasamatsu2004,Kasamatsu2006,Sabari2019}
BECs. 
As pointed out in~\cite{ZRaptiMI,Baizakov 2C MI,Ruostekoski}, 
the above scenario on MI studies can significantly change 
by considering discrete multi-component systems.

Beyond the GP mean-field formalism, by considering quantum fluctuations
through the Lee-Huang-Yang (LHY) term~\cite{1957LHY}, the MI was studied in 
Refs.~\cite{2019-Abdullaev,2022-Otajonov,2024-Otajonov}, following analysis of Faraday 
wave patterns and droplets generated in Bose gas mixtures~\cite{2005Abdullaev}.  
Flat bands and dynamical localization of binary BEC mixtures have also been 
studied in discrete media, in~\cite{2018Abdullaev}, where 
a deep optical lattice with periodic time modulation of the Zeeman field was assumed.
However, the onset conditions of MI in discrete media have not been 
{\it completely} explored particularly when assuming SO-coupled condensates.

In light of the above developments, it would be interesting to investigate the 
freedom associated with Rashba-type SOC in BECs, intra/inter-species interaction 
and dispersion. Such studies could lead to the identification of several 
stable domains of interest to be further investigated in possible experimental setups.
This is one of the main purposes of the present study that we are reporting, 
within an attempt to analyze the impact of SO coupling on MI in discrete BECs.

In addition, to explore the dynamics of SOC-BEC in an OL, we solve the two-component 
discrete GP equations using direct numerical integration. For this purpose, we use fourth-order
Runge–Kutta (RK4) algorithm. This scheme is widely used for nonlinear Schrödinger-type systems,
including GP equations. At each time step, this algorithm evaluates the nonlinear
interaction terms, nearest-neighbor hopping, and SOC contributions, thereby 
advancing the coupled wave functions in time. 

The structure of this paper is the following: In section~\ref{sec2}, we present the
tight-binding model, with the SO coupling model constructed in terms of coupled 
discrete GP equations.
In section~\ref{sec3}, we present the associated dispersion relations, obtained 
via linear stability analysis. 
In section~\ref{sec4}, the main modulational stability results are presented and
discussed for different signs of the intra- and inter-component fragmentations, 
with SO coupling interaction. 
Finally, in section~\ref{sec5}, we elaborate on the salient features of the 
present investigation.  

\section{The tight-binding model}
\label{sec2} 
For a discrete two-component spin-1/2 tight-binding model of Bose-Einstein Condensate (BEC), 
with SOC loaded in a one-dimensional (1D) deep optical lattice, the system of 
coupled equations can follow the formalism as presented in Refs.~\cite{2015-Salerno,2016-Salerno}.  
In these references, an inter-species SOC between the two hyperfine states was employed,
whereas in some other approaches, such as in Ref.~\cite{Sabari2021}, intra-species coupling 
has been assumed. 
Both approaches can be considered in experimental setups, with the 
intra-species SOC modifying the spin textures within single BECs;
and inter-species SOC introducing momentum-dependence coupling between 
two distinct BECs, leading to richer dynamics and novel quantum phases. 
The intra-species SOC is easy to be experimentally implemented, as it
occurs between hyperfine states of the same atomic species. It can be 
implemented by using Raman laser coupling to induce transitions between 
the states. See~\cite{2015Belicev} and references therein.
However, the inter-species SOC is experimentally more challenging 
due to the need for selective species addressing. It requires precise 
control of laser frequencies to avoid unwanted transitions and may involve 
interspecies Feshbach resonances to tune interactions, such that can lead 
to richer dynamics and novel quantum phases. 
See, for example, Refs.~\cite{2005Ruseckas,2012Edmonds}, as well as the discussion in Ref.~\cite{2016-Salerno}.
Therefore, 
for the linear part,  corresponding to the single-particle Hamiltonian, 
it was found instrutive to consider
the two main usual approaches for the Dresselhaus-Rashba SOC scheme (see also 
Ref.~\cite{2015Belicev} for details). The approaches are labeled as (A) and (B), where A refers to inter-species
SO coupling assumption, which we are going to use along the main part of the present work. The approach B,
which refers to the intra-species SO coupling, we are also presenting for completeness, as to evidence the 
main differences in the formalism.
With the addition of a confining periodic optical lattice (OL) potential in the 
$x-$direction, $V_{ol}(x)\equiv V_0 \cos(2k_L x)$ (with $k_L$ being the lattice 
wave-number), which can be generated by counter-propagating laser fields, 
the respective single-particle linear Hamiltonian operators, $H_A$ and $H_B$, 
for particle with mass $m$, are given by
\begin{eqnarray}
H_{A} &=& \frac{p_x^2}{2m}+V_{ol}(x)+ \frac{\hbar^2\kappa}{m}  p_x\sigma_x +\hbar \Omega_Z \sigma_z,\label{HA}\\
H_{B} &=& \frac{p_x^2}{2m}+V_{ol}(x)+ \frac{\hbar^2\kappa}{m} p_x\sigma_z +\hbar \Omega_Z \sigma_x,\label{HB}
\end{eqnarray}
where $\kappa$ and $\Omega_Z$ are, respectively, the SOC parameter and 
Zeeman-field frequency; with 
$\sigma_x = \left({\begin{array}{ccc} 0&1\\1&0\end{array}}\right)$ and
$\sigma_z =\left({\begin{array}{ccc} 1 & 0\\ 0&-1 \end{array}}\right)$ being
the usual spin$-\frac{1}{2}$ Pauli matrices.
From the three-dimensional (3D) GP formalism, by considering the transversal trap 
frequency $\omega_y=\omega_z\equiv\omega_\perp$ much larger than the longitudinal 
one, $\omega_\perp\gg\omega_{x}=\omega_\parallel$, we obtain a cigar-type quasi-1D trap model
for the BEC mixture.
In the presence of SOC,  the corresponding  formalism can be written in matrix form as
{\small
\begin{eqnarray}\label{eqGP0}
{\rm i}\hbar\frac{\partial\Psi}{\partial t} & =& \left[
H_{A,B}+H_{nl}\right]\Psi ,
\label{HNL}\\
H_{nl}&\equiv& 2\hbar\omega_\perp\left(
\begin{array}{cc}
\sum_j a_{1j}|\psi_j|^2 & 0\\
0&\sum_ja_{j2}|\psi_j|^2\\
\end{array}
\right)
 \nonumber
,\end{eqnarray}
}where $\Psi\equiv\Psi(x,t)$ 
$\equiv \left( \begin{array}{c} \psi_1(x,t)\\ \psi_2(x,t) \end{array} \right)$
$\equiv \left( \begin{array}{c} \psi_1\\ \psi_2 \end{array} \right)$
is the two-component total wave function normalized to the total number 
of atoms, $N=\sum_{j=1}^2 \int dx |\psi_j|^2,$ with
$a_{jj}\;\;(j=1,2)$ and $a_{12}$ being the two-body scattering lengths between
intra- and inter-species of atoms.
As noticed from Eqs.~\eqref{HA} and \eqref{HB}, the formalisms for the two models,  A and B, 
only differ by the interchange of the linear Dresselhaus-Rashba couplings.

To reach dimensionless equations, the space-time variables are redefined, as follows:
$x\to{x}/{k_L}$ and $t \to {t}/{\omega_R}$, where $\omega_R\equiv {E_R}/{\hbar}
\equiv {\hbar k_L^2}/{(2m)}$. In this case, we can write the OL potential and 
wave-function components as 
\begin{eqnarray}
&&{V_{ol}(x)} \to E_R V(x),\;\;{\rm with}\;\;V(x)\equiv V_0\cos(2x),\nonumber\\  
&&\psi_j \equiv \sqrt{\frac{\omega_R}{2\omega_{\perp}a_0}}  \psi_j(x,t).
\label{ol-wf}\end{eqnarray}
In these definitions, $E_R$ is the recoil energy with $a_0$ being the background 
scattering length. 
We further define the parameters as in \cite{2016-Salerno}, with
\begin{eqnarray}
b&\equiv& \frac{2\kappa}{k_L},\;\; \nu\equiv\frac{\Omega_Z}{\omega_R},\;\;
\tilde{g}=\frac{a_{jj}}{a_0},\;\; \tilde{g}_{12}=\frac{a_{12}}{a_0},
\label{g-wf}\end{eqnarray}
to write down the GP formalisms, for the two models, in dimensionless form. 
With $j=1,2$ labeling the two components, for model A (inter-SOC), we have
\begin{eqnarray}
{\rm i} \frac{\partial \psi_{j}}{\partial t} & =& \left(-\frac{\partial^2}{\partial x^2} + V(x) 
-(-1)^{j}  \nu \right) \psi_{j}
- {\rm i} b \frac{\partial \psi_{3-j}}{\partial x} + \nonumber \\
&& \left(\tilde{g} |\psi_{j}|^2+ {\tilde{g}_{12}} |\psi_{3-j}|^2\right) \psi_j
;\label{SO-GPA}
\end{eqnarray}
whereas, for model B  (intra-SOC),
\begin{eqnarray}
{\rm i} \frac{\partial \psi_{j}}{\partial t} & =& \left(-\frac{\partial^2}{\partial x^2} + V(x) 
+(-1)^{j}{\rm  i} b \frac{\partial}{\partial x}\right) \psi_{j}
+\nu\psi_{3-j}  + \nonumber \\
&& \left(\tilde{g} |\psi_{j}|^2+ {\tilde{g}_{12}} |\psi_{3-j}|^2\right) \psi_{j} 
.\label{SO-GPB}
\end{eqnarray}
With definitions \eqref{ol-wf} and \eqref{g-wf}, the normalization 
of the total number of atoms $N$ can be written as
\begin{equation}
{N}= 
\frac{\omega_R}{2\omega_{\perp}k_La_0}
\sum_{j=1}^2\int dx|\psi_j|^2=
\frac{\omega_R}{2\omega_{\perp}k_La_0}\overline{N},
\end{equation}
with $\overline{N}$ being the corresponding rescaled number.

Once represented the two possible SOC approaches by \eqref{HA} (inter-SOC) and \eqref{HB} 
(intra-SOC), and noticing that model B has already been investigated in the same context
in Ref.~\cite{Sabari2021}, we follow the present investigation by studying modulational instability
considering the tight-binding model applied to the inter-SOC (model A). 
Both coupling models have particular characteristics, whose advantages can be 
experimentally exploited. 

Within the tight-binding approximation, the OL is generated 
by the periodic potential defined in \eqref{ol-wf}, with $V_0\gg 1$, for
both components, by first considering the uncoupled linear periodic 
eigenvalue problem ($\Omega_Z=0$, $\kappa=0$, $g=g_{12}=0$), in which the two-component 
wave functions $\psi_j$ are expanded in terms of the Wannier functions 
$w_m(x)$~\cite{1937-Wannier},
as introduced in~\cite{2002-Alfimov}.
Here, we follow closely Refs.~\cite{2015-Salerno,2016-Salerno},
with the time-dependent expansion coefficients, $\phi^\pm_n\equiv\phi^\pm_n(t)$,
carrying the symbols $+$ and $-$ that are associated, respectively, with 
components 1 and 2 of the wave function $\psi_j$. So, for the inter-SOC formalism
given in Eq.~\eqref{SO-GPA}, considering the wave functions 
\begin{eqnarray}
\psi_1\equiv\sum_{n,m}\phi^+_n w_m(x-n), 
\psi_2\equiv\sum_{n,m}\phi^-_n w_m(x-n),
\label{wfexpand}    
\end{eqnarray}
we obtain
{\small\begin{eqnarray}
{\rm i}\frac{d \phi^{\pm}_n}{dt}&=&
-\Gamma(\phi^{\pm}_{n+1}+\phi^{\pm}_{n-1})+ {\rm i}\gamma (\phi^{\mp}_{n+1}-\phi^{\mp}_{n-1})\nonumber\\
&\pm& \nu \phi^{\pm}_n+\left(g |\phi^{\pm}_n|^2+ g_{12}  
|\phi^{\mp}_n|^2\right) \phi^{\pm}_n ,
\label{dis1d}
\end{eqnarray}
}where $\Gamma$ is the hopping coefficient of adjacent lattice sites,
{\small
\begin{eqnarray}
\Gamma &\equiv& \Gamma_{n,n+1}=\int w^*(x-n)\frac{\partial^2}{\partial x^2}w(x-n-1) dx,
\label{Gamma}
\end{eqnarray}
}while $g$ and $g_{12}$ are the intra- and inter-component collision strengths,
respectively expressed in terms of integrals on the Wannier functions by
{\small \begin{eqnarray}
g &=& {\tilde{g}} \int |w(x-n)|^4 dx,\;
g_{12} = {\tilde{g}_{12}} \int |w(x-n)|^4 dx,\\
\gamma&\equiv& \gamma_{n,n+1}= b \int w^*(x-n)\frac{\partial}{\partial x} w(x-n-1) dx.
\end{eqnarray}
}Here, it is appropriate to remark that due to the strong localization of the 
Wannier functions around the lattice sites, the sums on $n$ have been restricted 
to on-site and next neighborhood sites only, for both diagonal and non-diagonal
terms, within a model also being considered in Refs.~\cite{2015Belicev,2015-Salerno}.
Also due to the properties of Wannier functions, in the derivation of \eqref{dis1d} 
the following relations were applied among the coefficients: 
$\Gamma_{n,n+1}=\Gamma_{n,n-1}$, 
$\gamma_{n,n}=0$, 
$\gamma_{n,n-1}=-\gamma_{n-1,n}=-\gamma_{n,n+1}$. 
The rescaled total number of atoms can be written as
$\overline{N}= \sum_{n,\pm} |\phi^{\pm}_n|^2.$
Within the same above notation and coefficient definitions, 
for the intra-SOC case (defining the discretized components by
$\hat\phi_n^{\pm}$), the equation
corresponding to \eqref{dis1d} can be written as 
{\small\begin{eqnarray}
{\rm i}\frac{d \hat\phi_n^{\pm}}{dt}&=&
-\Gamma(\hat\phi^{\pm}_{n+1}+\hat\phi^{\pm}_{n-1})\mp {\rm i}\gamma (\hat\phi^{\pm}_{n+1}-\hat\phi^{\pm}_{n-1})\nonumber\\
&+& \nu \hat\phi^{\mp}_n+\left(g |\hat\phi^{\pm}_n|^2+ g_{12}  
|\hat\phi^{\mp}_n|^2\right) \hat\phi^{\pm}_n .
\label{dis1dB}
\end{eqnarray}
By comparing \eqref{dis1d} with \eqref{dis1dB} we can appreciate the exchange role 
of the spin-orbit parameter $\gamma$ and the
Zeeman field parameter $\nu$ (also called Rabi parameter). Next, we will 
concentrate the linear stability analysis on the inter-SOC as given by \eqref{dis1d}.
}
\section{Linear stability analysis}
\label{sec3}
The stability of the solutions is going to be explored by considering the corresponding
discretization, with the $n$ sites of both $\pm $ components represented by the
ansatz plane-wave solutions,
\begin{equation}
\phi_{n}^{\pm }=u_{\pm }\exp [{\rm i}\left(\omega t + n q\right)],
\label{ansat}
\end{equation}
where it is being assumed that the oscillating frequencies $\omega$ and wave numbers 
$q$ must be the same for both $\pm $ components, in view of their couplings. Also, both
$\pm$ component amplitudes $u_\pm$ are taken as the same for each site $n$.
The above $\omega$, apart of a constant shift $2\Gamma$ ($\omega\to\omega-2\Gamma$), 
has the following dependence on the wave-number $q$ and linear couplings, $\nu$ and $\gamma$:
{\small
\begin{eqnarray}
\left[\omega-2\Gamma\cos(q)\mp\nu+(g u_\pm^2+g_{12}u_\mp^2)\right]u_\pm=
{2\gamma\sin(q)}{u_\mp}\label{spec-coupl}
.\end{eqnarray}
}From the above, the corresponding linear relation is obtained with $g=g_{12}=0$. 
In this limiting case (with  $\omega\to\omega_0$), the two branch spectral 
solutions are given by
\begin{eqnarray}
\omega_{0\pm}&=&2\Gamma\cos(q)
\pm\sqrt{\nu^2+[{2\gamma\sin (q)}]^2}
.\label{linearsol}
\end{eqnarray}
Next, we assume that the solutions provided by \eqref{ansat} are 
slightly perturbed, at each discrete site $n$, by an oscillating 
time-dependent complex term $\xi_n^{\pm}\equiv\xi_n^{\pm}
(t;\Omega,Q)$,
{\small 
\begin{eqnarray}
\xi_{n}^{\pm} \equiv \chi^{\pm}_{a}\cos(\Omega t+nQ)+{\rm i}\chi^{\pm}_{b}\sin(\Omega t + nQ)
{\rm e}^{-{\rm i}(\Omega t +nQ)},
\label{xis}
\end{eqnarray}
}having absolute values smaller than $|u_\pm|$, such
that only first-order terms are retained in the formalism 
for the inter-SOC \eqref{dis1d}. So, 
the unperturbed $\phi_{n}^\pm$ are replaced by 
\begin{eqnarray}
{\bm\phi}_{n}^{\pm }&=&(u_{\pm }+\xi_n^{\pm})
e^{{\rm i}( \omega t+ n q)},\label{Phi-pert}
\end{eqnarray}
which is followed by the linearization of the nonlinear
terms. Within this procedure, a discrete differential equation 
for $\xi_n^\pm$ is obtained, similar as in 
Refs.~\cite{2015Belicev,Sabari2021}. 
The corresponding perturbed equations for $\xi^{\pm}$ are: 
{\small\begin{eqnarray}
{\rm i}\frac{\partial \xi^\pm_n}{\partial t}\!\!&=&\!\!
\left[-\Gamma(\xi^\pm_{n-1}+\xi^\pm_{n+1}-2\xi^\pm_n)
+{\rm i}\gamma(\xi^\mp_{n+1}-\xi^\mp_{n-1})\right]\cos(q)\nonumber \\
&-&\!\!\!\left[\gamma\left(\xi^\mp_{n-1}+\xi^\mp_{n+1}
-\frac{2 u_\mp\xi^\pm_n}{u^\pm}\right)\!\!
+{\rm i}\Gamma(\xi^\pm_{n+1}-\xi^\pm_{n-1})
\right]\sin (q) \nonumber \\
&+&g u_\pm^2(\xi^\pm_n+\xi^{\pm *}_n)
+g_{12} u_+u_-(\xi^\mp_n+\xi^{\mp *}_n)\label{pert-xin}.
\end{eqnarray}
}
Quite relevant is to verify that, within the above coupled
equation for the perturbation $\xi_n^\pm$, there is no explicit 
dependence on the Rabi parameter $\nu$. This points out 
the main difference from the intra-SOC case.
Apart from this fact, the inter-SOC dispersion relation has 
the same format as in the case of intra-SOC. The perturbation wave
parameter $\Omega$, introduced in \eqref{xis}, is given by the solutions 
of the following quartic polynomial expression:
\begin{align}
\begin{split}
\Omega^4+P_3\Omega^3+P_2\Omega^2+P_1\Omega+P_0=0,  \label{equ9}
\end{split}
\end{align}
\noindent where the coefficients $P_{j=0,1,2,3}$ are 
given in terms of the wave-numbers $q$, $Q$, and parameters 
$\Gamma$, $\gamma$, $g$, $g_{12}$.
By defining
$\Gamma_\pm\equiv 2\Gamma\,[\cos(q)-\cos\left(q\pm Q\right)]=4\Gamma\sin\left(\frac{Q}{2}\right)\sin\left(\frac{Q}{2}\pm q\right),$
$\gamma_0\equiv 2\gamma\sin(q)$, $\gamma_\pm\equiv 2\gamma\sin(q\pm Q)=
\gamma_0\pm 4\gamma\sin\left(\frac{Q}{2}\right)\cos\left(\frac{Q}{2}\pm q\right),$ we have
{\small\begin{eqnarray}
P_3&=&2(\Gamma_+-\Gamma_-), \nonumber\\
P_2&=&\Gamma^2_++\Gamma_-^2-4 \Gamma_- \Gamma_+ -2 \gamma_0(\Gamma_+ + \Gamma_- +\gamma_0)-\gamma_-^2 - \gamma_+^2 \nonumber\\
&-&2g \left( \Gamma_+ +\Gamma_- + 2\gamma_0\right)+  2 g_{12} \left(\gamma_+ + \gamma_- \right),\nonumber\\
P_1&=& 2\left[\Gamma_+\Gamma_-+\gamma_0^2+\gamma_0(\Gamma_++\Gamma_-)\right](\Gamma_--\Gamma_+)\nonumber\\
&+&2(\gamma_+^2\Gamma_--\gamma_-^2\Gamma_+)+2\gamma_0(\gamma_+^2-\gamma_-^2)\nonumber\\
&+&2g\left[(\Gamma_--\Gamma_+)(\Gamma_-+\Gamma_++2\gamma_0)+\gamma_+^2-\gamma_-^2\right] \nonumber\\
&+&4g_{12}\left[\Gamma_+\gamma_- - \Gamma_-\gamma_+ +\gamma_0(\gamma_- - \gamma_+)\right],  \nonumber\\
P_0&=&\Gamma_-^2 \Gamma_+^2 +2 \gamma_0 [\Gamma_- (\Gamma_+^2-\gamma_+^2)
+\Gamma_+(\Gamma_-^2-\gamma_-^2)] \label{Ps}\\
&+&\gamma_0^2[(\Gamma_-^2-\gamma_-^2)+(\Gamma_+^2-\gamma_+^2)+4\Gamma_- \Gamma_+] \nonumber\\
&+&2 \gamma_0^3 (\Gamma_-+\Gamma_+) + \gamma_0^4-\gamma_+^2 \Gamma_-^2-\gamma_-^2 \Gamma_+^2+\gamma_-^2 
\gamma_+^2 \nonumber\\
&+&(g^2-g_{12}^2)[(\Gamma_- + \Gamma_++2\gamma_0)^2 
- (\gamma_- + \gamma_+)^2]+ 6g\gamma_0^2(\Gamma_- +\Gamma_+) \nonumber\\
&+&2g \gamma_0 [(\Gamma_- +\Gamma_+)^2+2 \Gamma_- \Gamma_+
+2\gamma_0^2-(\gamma_-^2 + \gamma_+^2)]\nonumber\\
&+&2 g[(\Gamma_- (\Gamma_+^2-\gamma_+^2)+\Gamma_+(\Gamma_-^2-\gamma_-^2)] \nonumber\\
&+&4g_{12}\gamma_0 (\gamma_+ \Gamma_-+\gamma_- \Gamma_+)
+2g_{12}(\gamma_- \Gamma_+^2 +\gamma_+ \Gamma_-^2) \nonumber\\
&+&2 g_{12}(\gamma_- +\gamma_+) (\gamma_0^2 - \gamma_- \gamma_+)\nonumber
.\label{equ10}\end{eqnarray}
}

For the particular two kind of grid arrangements, unstaggered (with $q=0$, when there is no change 
in the site positions) and staggered (with $q=\pi$), 
we have $\gamma_0=0$, with 
$\Gamma_\pm \equiv (1-2\delta_{q,\pi})\widetilde\Gamma$,
$\gamma_\pm= \mp (1-2\delta_{q,\pi}) \widetilde\gamma$,
where $\widetilde\Gamma\equiv 2\Gamma[1-\cos(Q)]$ and $\widetilde\gamma\equiv 
2\gamma\sin(Q)$ (with $\delta_{q,\pi}=1$
for $q=\pi$ and zero otherwise).
In both cases, the coefficients have the same 
formal simplified expression. For the staggered mode ($q=\pi$), they are
given by
{\small\begin{eqnarray}
P_3&=&0,\nonumber\\
P_2&=&-2\left[\Gamma_+(\Gamma_+ +2g)+ \gamma_+^2\right],\nonumber\\
P_1&=&-8g_{12}\Gamma_+\gamma_+
,\label{q=pi}\\
P_0&=&[\Gamma_+(\Gamma_+ +2g)-\gamma_+^2]^2-
4g_{12}^2\Gamma_+^2 
.\nonumber
\end{eqnarray}
}So, when $\gamma=0$ (no SOC) or $g_{12}=0$ (without nonlinear coupling) 
($P_3=P_1=0$), these expressions provide straightforward solutions 
for \eqref{equ9}.
In the limiting case with $\gamma=0$, in the staggered 
mode ($q=\pi$), the four possible solutions of $\Omega$ are given by
as {\footnotesize
\begin{eqnarray}
\Omega^2_\pm &=& \Gamma_+^2+2(g\pm g_{12})\Gamma_+\nonumber\\
&=& 4[\Gamma(\cos(Q)-1)]\left[\Gamma(\cos(Q)-1)+(g\pm g_{12})\right]\nonumber\\
\Omega_\pm&=&2
\sqrt{\Gamma(\cos Q -1)\left[\Gamma(\cos Q-1)+(g\pm g_{12})\right]}
\nonumber\\
&=&4\Gamma\sin\left(\frac{Q}{2}\right)
\sqrt{\sin^2\left(\frac{Q}{2}\right)-\frac{g\pm g_{12}}{2\Gamma}}
\label{g120qpi}.\end{eqnarray}
} The overall negative $\Omega$ solutions are ignored, because 
the perturbation parameter solutions affecting the MI are verified by
the absolute values of possible nonzero imaginary parts of $\Omega_\pm$. 
So, in this particular case with $\gamma=0$, for all the staggered s
olutions ($q=\pi$) of \eqref{Phi-pert}, the growth rate of the 
instability is affected by the non-zero values of 
\begin{eqnarray} \zeta_\pm&\equiv &2 |\text{Im} (\Omega_\pm)|\label{MI}
\\
&=&\text{Im}\left|8\Gamma\sin\left(\frac{Q}{2}\right)
\sqrt{\frac{g\pm g_{12}}{2\Gamma}-{\sin^2\left(\frac{Q}{2}\right)}}\right|
\nonumber.
\end{eqnarray}
Otherwise, the $\bm\phi_\pm$ solutions will be stable.
As noticed, the two possible non-zero $\zeta_\pm$ solutions differ 
only by the interspecies interaction $g_{12}$.
However, in the above it was considered $P_1=0$, such that 
for the MI results having some dependence on the SO parameter $\gamma$ 
is essential that $g_{12}\ne 0$, when the simplified expression
\eqref{g120qpi} is no longer valid.
So, these expressions are useful as guides for the 
more general numerical solutions, which can happen (in the 
staggered mode) when both $g_{12}$ and $\gamma$ are nonzero. 

As already pointed out when deriving the discrete perturbed coupled expression
\eqref{pert-xin},
differently from the intra-SOC case (with the corresponding dispersion 
relation provided in \cite{Sabari2021}), one should notice that the 
oscillating frequency $\Omega$ of the perturbation is not affected by 
the Zeeman field parameter $\nu$ in the inter-SOC case.
As shown by the linear spectrum \eqref{linearsol}, together with
\eqref{spec-coupl}, $\nu$ splits the two branch solutions, 
but is not coupling the solutions, being canceled out when 
considering the perturbations, implying that the MI results
depend only on the SOC and nonlinear parameters.
Our main results for the modulational instability are provided 
in the next section. With mostly of the results being related to the inter-SOC 
case (model A), a sample case is provided for the intra-SOC, as to be
compared with previously obtained results given in
\cite{Sabari2021}.

\section{Modulational Stability Results}
\label{sec4}
Through the four possible solutions $\Omega_i (i=1,2,3,4)$ of Eq.~(\ref{equ9}),
considering the perturbations defined in \eqref{xis}, the growth rate 
instabilities are provided by the corresponding imaginary parts, called gains,
which are defined as $\zeta_i=2 |\text{Im} (\Omega_i)|$. 
A few significant results are given in the following, in which the
instabilities are presented in diagrammatic planes parametrized by the 
wave numbers ($q$ and $Q$), the nonlinear interactions ($g$ and $g_{12}$),
as well as the parameter $\gamma$ for the linear SO coupling. 
The hoping (kinetic) coefficient is fixed to $\Gamma=1$ 
in this study.

\begin{figure}[!ht]
\includegraphics[width=.22\textwidth]{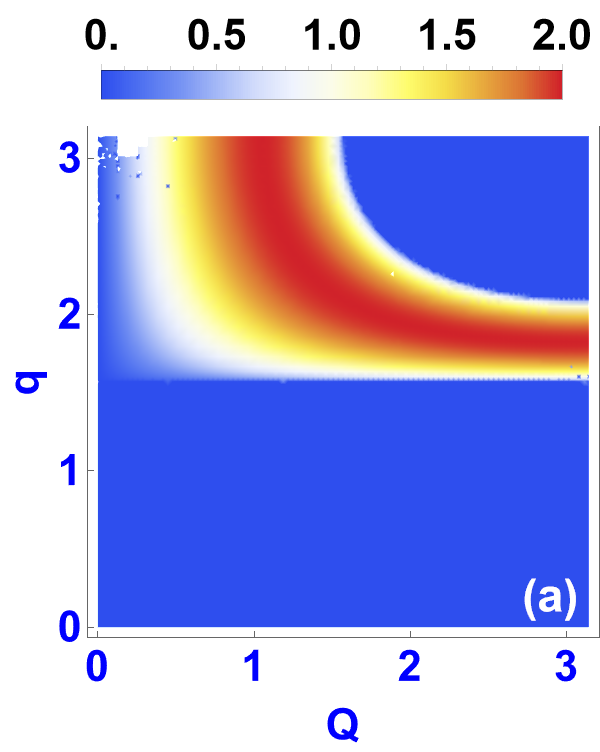} 
\includegraphics[width=.22\textwidth]{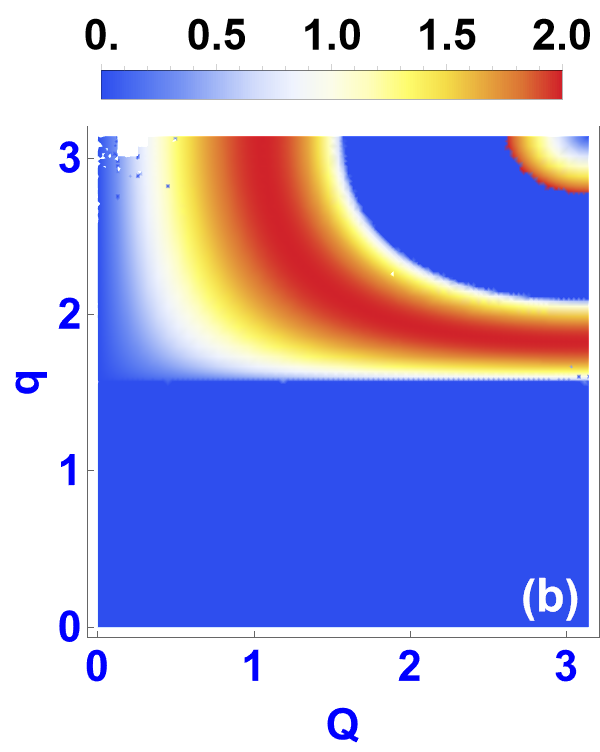}
\caption{(color online) 
Inter-SOC solutions obtained from Eq.~\eqref{Phi-pert},
shown in diagrams of $q$ versus $Q$,
with MI growth rates (indicated by the color bars) given by 
non-zero $2|{\rm Im}(\Omega)|$, obtained from \eqref{equ9}. 
In this particular case with $\gamma=0$, 
the other parameters are $\Gamma=1$, $g=1$, and $g_{12}=1$. 
Both solutions refer to $\zeta_+$, as noticed in the
limit $q=\pi$.
Within the dimensions given in the text, all quantities 
are dimensionless.
}
\label{fig01}
\end{figure}
\begin{figure*}[!ht]
\begin{center}
\includegraphics[width=0.99\textwidth]{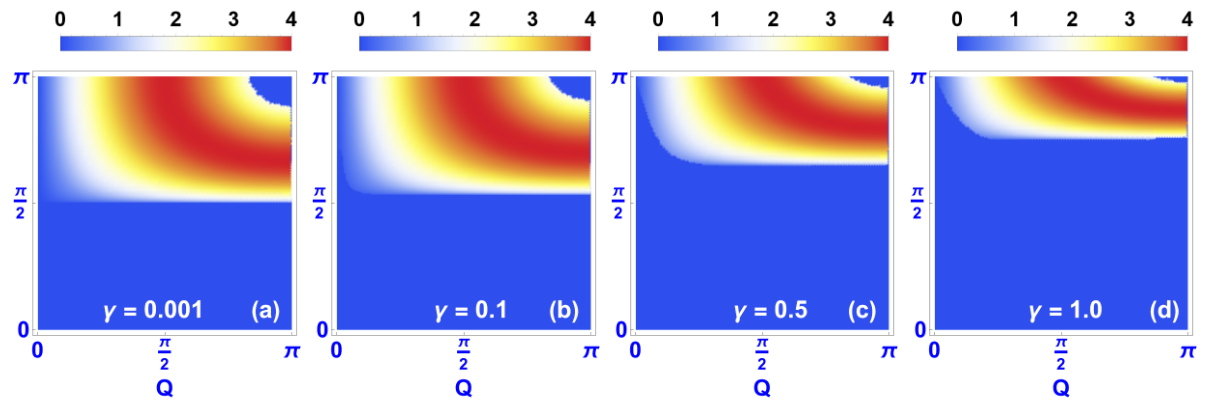}
\end{center}
\caption{(color online) 
Effect of SOC on the inter-SOC solution obtained with growth rate $\zeta_+$.
The growth rates are indicated by the color bars.
The results are shown as functions of $q$ versus $Q$, considering the given
values of $\gamma=0.001 (a), 0.1 (b), 0.5$ (c) and $1.0$ (d). The remaining parameters are 
$g=1$, $g_{12}=1$, and $\Gamma=1$.
All quantities are dimensionless, with dimensions as defined in the text.}
\label{fig02}
\end{figure*}
\begin{figure*}[!ht]
\begin{center}
\includegraphics[width=1.0\textwidth]{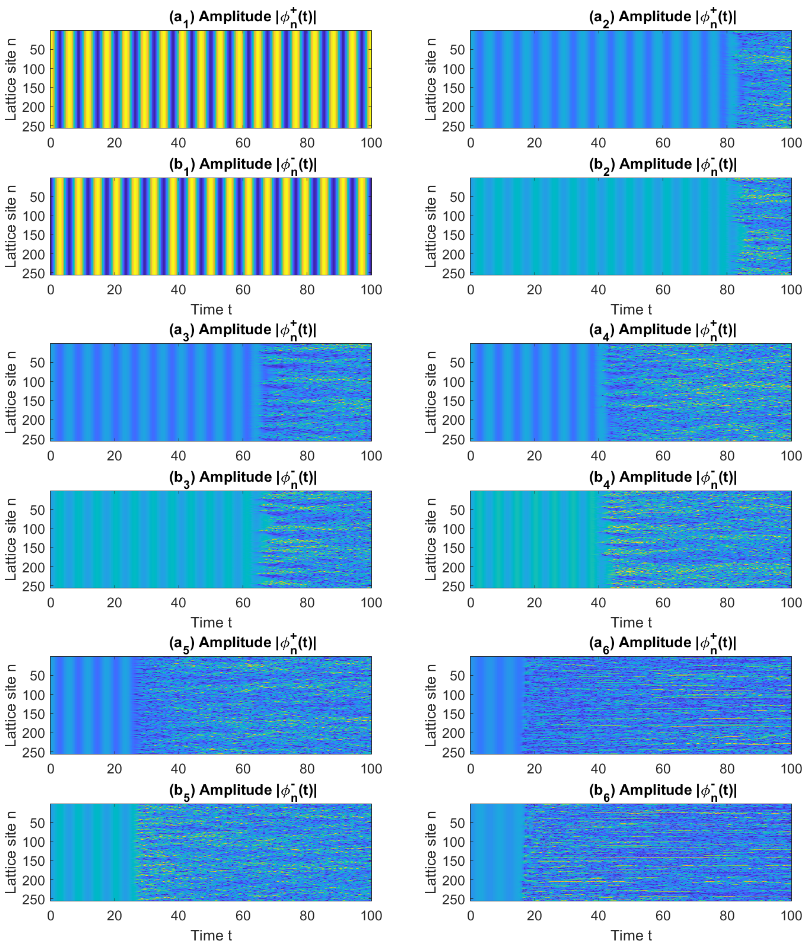}
\end{center}
\caption{(Color online) 
Spatiotemporal evolution of the wave amplitudes \( |\phi_n^{+}(t)| \) (top rows) and \( |\phi_n^{-}(t)| \) (bottom rows) for different carrier wave numbers \( q \). Parameters are fixed at \( \Gamma = g = g_{12} = 1 \), and SO coupling strength \( \gamma = 0.1 \). The initial condition is a plane wave perturbed by a long-wavelength modulation. The figures illustrate the transition from stability to MI as \( q \) increases. The evolution of the wave amplitude is shown for carrier wave numbers \( q = 1.0, 1.6, 1.8, 2.0, 2.5, 3.0 \) (panels $a_{1-6}, \, b_{1-6}$ respectively)}
\label{fig02_amp}
\end{figure*}

\begin{figure*}[!ht]
\begin{center}
\includegraphics[width=1.0\textwidth]{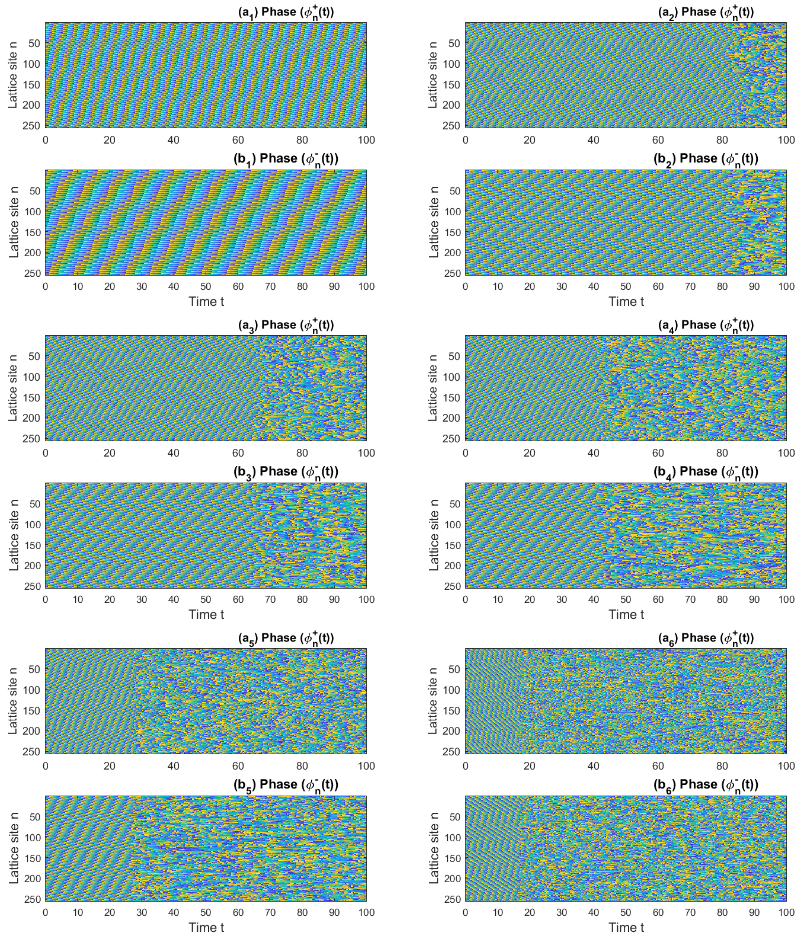}
\end{center}
\caption{(color online) Spatiotemporal evolution of the phases \( \arg[\phi_n^{+}(t)] \) (top rows) and \( \arg[\phi_n^{-}(t)] \) (bottom rows) corresponding to the amplitudes in Fig.~\ref{fig02_amp}. Parameters are identical: \( \Gamma = g = g_{12} = 1 \), and spin-orbit coupling strength \( \gamma = 0.1 \). The evolution of the phase is shown for carrier wave numbers \( q = 1.0, 1.6, 1.8, 2.0, 2.5, 3.0 \) (panels $a_{1-6}, \, b_{1-6}$ respectively)}
\label{fig02_pha}
\end{figure*}
\begin{figure*}[!ht]
\begin{center}
\includegraphics[width=1.0\textwidth]{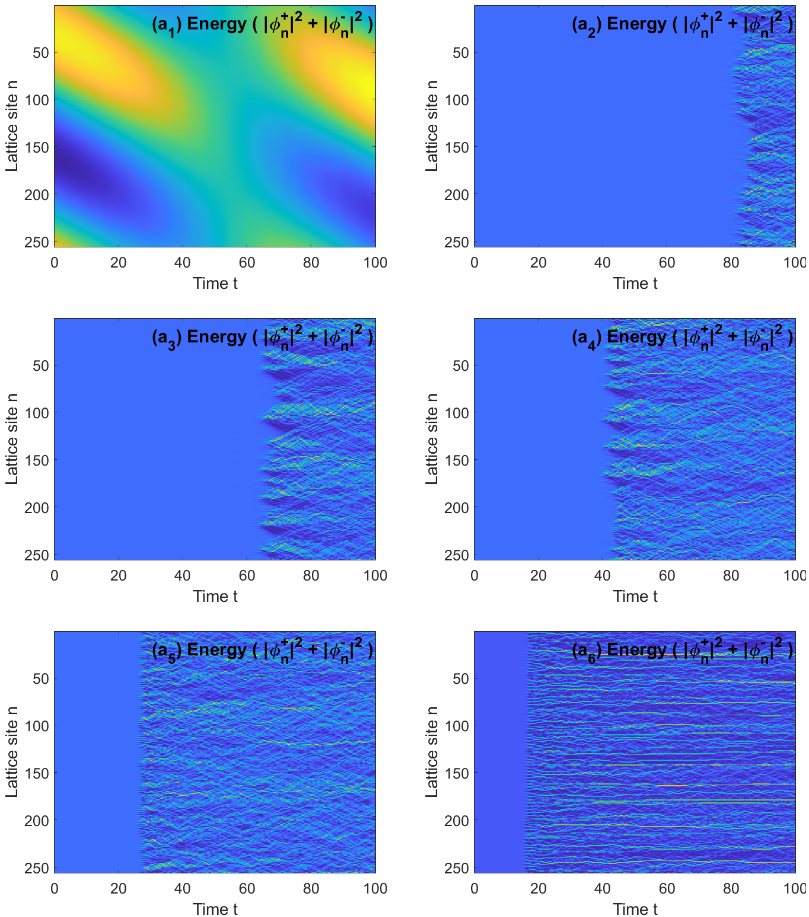}
\end{center}
\caption{(Color online) Spatiotemporal evolution of the total on-site energy density $\mathcal{E}_n(t) = |\phi_n^+(t)|^2 + |\phi_n^-(t)|^2$ 
for carrier wave numbers $q = 1.0, 1.6, 1.8, 2.0, 2.5, 3.0$ (panels $a_1-a_6$ respectively). 
All other parameters are fixed at $\Gamma = g = g_{12} = 1$, and $\gamma = 0.1$.}
\label{fig02_ene}
\end{figure*}

\begin{figure*}[!ht]
\begin{center}
\includegraphics[width=1.0\textwidth]{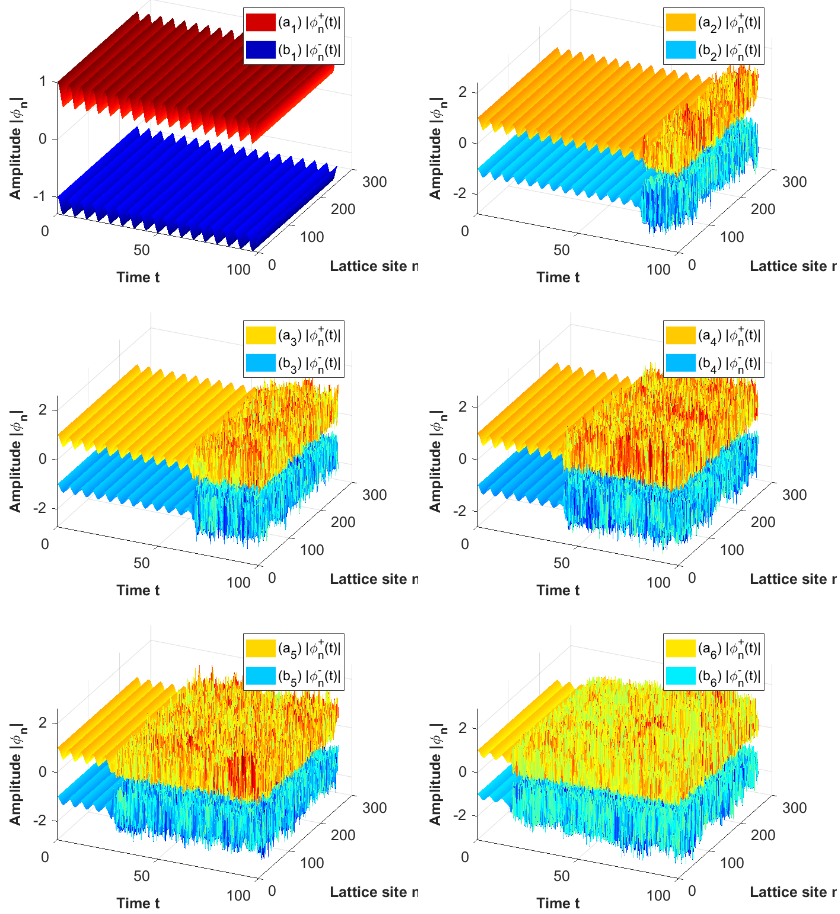}
\end{center}
\caption{(color online) Three-dimensional spatiotemporal evolution of the wave amplitudes 
for carrier wave numbers \( q = 1.0, 1.6, 1.8, 2.0, 2.5, 3.0 \) ($a_{1-6}, \, b_{1-6}$ respectively).
 Parameters are fixed at \( \Gamma = g = g_{12} = 1 \), and \( \gamma = 0.1 \). 
 These plots provide a complementary volumetric view of the MI dynamics shown in Figs.~\ref{fig02_amp}. The yellow and blue surfaces correspond to $|\phi_n^{+}(t)|$ and $|\phi_n^{-}(t)|$, respectively.}
\label{fig02_3d}
\end{figure*}

\begin{figure}[!ht]
\includegraphics[width=.2\textwidth]{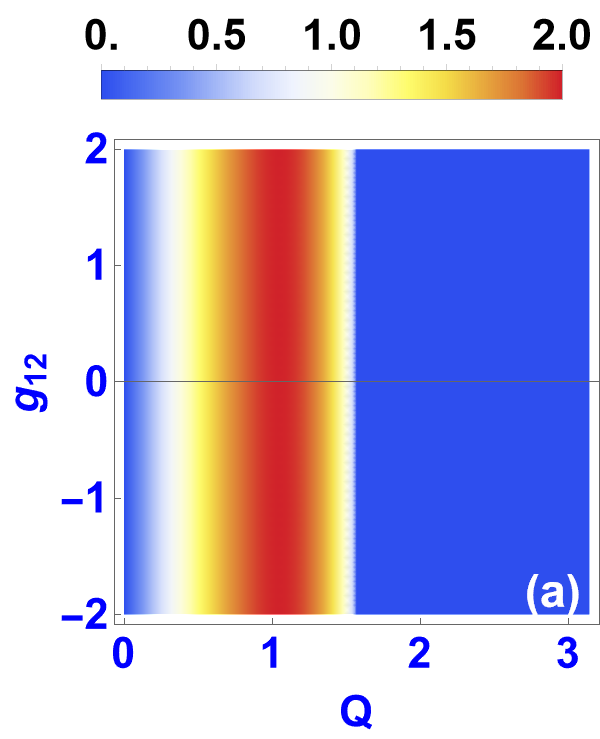} %
\includegraphics[width=.2\textwidth]{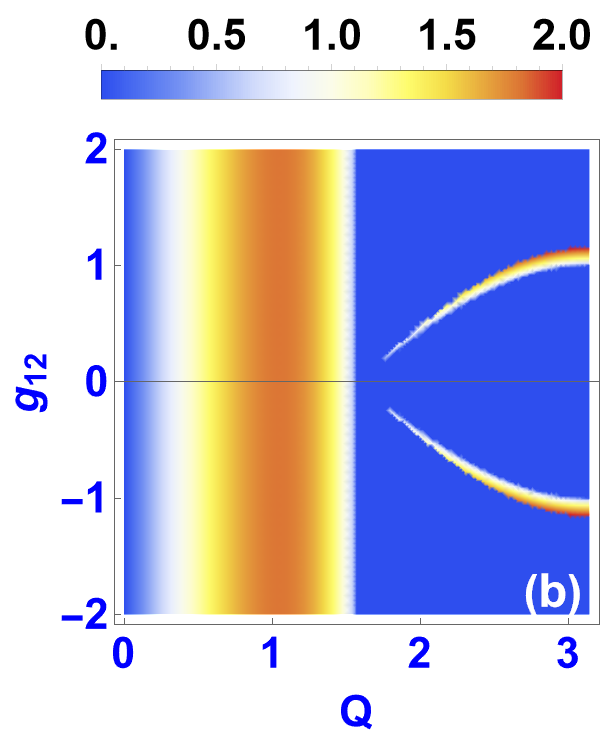}
\par
\includegraphics[width=.2\textwidth]{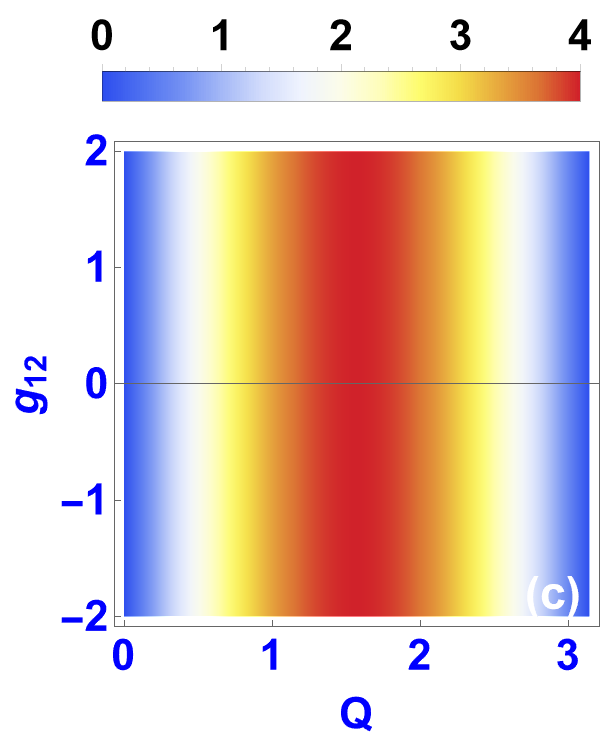} %
\includegraphics[width=.2\textwidth]{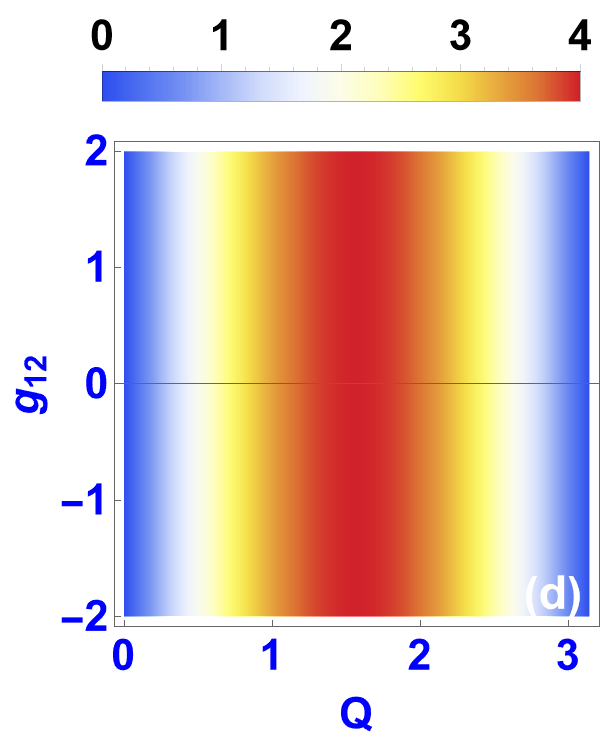}
\caption{(color online) 
The MI gains (with growth rates being indicated by color bars) are shown 
as functions of $Q$ vs $g_{12}$ and for $\gamma=0$ in the staggered mode ($q=\pi$). 
Left and right columns are for $\zeta_1$ and $\zeta_2$, respectively. 
The top and bottom rows are for $g=1$ and $g=2$, respectively. 
}
\label{fig03}
\end{figure}
Figure~\ref{fig01} provides two panels in diagrams of $q$ versus $Q$
illustrating the  
inter-SOC solutions for the particular case with $\gamma=0$. The solutions are
obtained from Eq.~\eqref{Phi-pert}, considering
MI growth rates given by the non-zero solutions of $2|{\rm Im}(\Omega)|$
[obtained from \eqref{equ9}]. 
In this case, the other parameters are $\Gamma=1$, $g=1$, and $g_{12}=1$,
such that the solutions refer to $\zeta_+$, as noticed in the
limit $q=\pi$.
In this case, there is no dependence on the linear couplings, as the MI 
is not affected by the Zeeman coupling, such that for the coupling only 
the nonlinear parameters $g$ and $g_{12}$ are relevant. 

{Figure~\ref{fig02} illustrates the effect of the SOC strength $\gamma$ on 
the inter-SOC solution, where the growth rate $\zeta_+$ is plotted as a function of $q$ verses $Q$. 
The analysis is performed for fixed interaction parameters $g=1$, $g_{12}=1$,
 and $\Gamma=1$, while systematically varying $\gamma$ within the range 
 $0 \leq \gamma \leq 1$. The corresponding growth rates are represented 
 by the color scale. The panels show the results for $\gamma=0.001$ (a), 
 $0.1$ (b), $0.5$ (c), and $1.0$ (d).
It is important to note that the instability sets in when $q \geq \pi/2$,
 with the unstable region progressively shifting towards $q=\pi$. 
 Furthermore, the extent of the instability region increases in both
  the $q$ and $Q$ directions as $\gamma$ grows.

Next, the analytical stability chart in Fig.~\ref{fig02}(b) is 
confirmed by direct numerical integration of the governing equations.
 Figure~\ref{fig02_amp} explicitly shows the emergence of MI from a 
 stable background as the carrier wave number \( q \) is varied from 1.01.0 to 3.03.0.
For \( q = 1.0 \) (Fig.~\ref{fig02_amp}$a_{1}, \, b_{1}$), both 
components exhibit persistent periodic modulations without any 
visible growth of localized perturbations or irregular fluctuations.
 This indicates that the system remains dynamically stable under 
 the chosen parameter regime, i.e., MI is absent in this case. 
 The initial perturbation disperses linearly without triggering
  significant nonlinear energy localization, indicating this 
  wave number resides within a stable region of the system's band structure.
For \( q = 1.6 \) to \( 2.5 \) (Figs.~\ref{fig02_amp} $a_{2-5}, \, b_{2-5}$),
 a clear onset and progression of MI is observed. The initial 
 uniform background becomes unstable and breaks into a train of coherent,
  localized structures. The number of wave packets increases
   and their spacing becomes more regular with increasing \( q \), 
   consistent with the predicted most-unstable modulation wave number from linear stability analysis.
For \( q = 3.0 \) (Fig.~\ref{fig02_amp} $a_{6}, \, b_{6}$), 
the instability develops more rapidly and results in a higher
 density of localized excitations. The complex interplay between 
 the nonlinearity, SOC, and the lattice dispersion is evident in the intricate patterns that emerge.
The symmetric development of instability in both components \( \phi_n^\pm \) is 
due to the balanced intra- and inter-species nonlinearities (\(g = g_{12}\)) 
and the specific form of the SOC.

The phase dynamics in Fig.~\ref{fig02_pha} provide complementary 
information to the amplitude evolution, revealing the coherence 
and velocity fields of the emerging wave packets.
For the stable case at \( q = 1.0 \) (Fig.~\ref{fig02_pha}$a_{1}, \, b_{1}$), 
the phase evolves in a nearly uniform and linear manner across the lattice, 
consistent with a stable plane wave whose phase velocity is constant in time and space.
As \( q \) increases within the MI regime (Figs.~\ref{fig02_pha}$a_{2-6}, \, b_{2-6}$), 
the phase profiles become increasingly disordered. The development of MI is 
marked by the formation of distinct phase slips and vortices 
(evident as discontinuous jumps in color from \( -\pi \) to \( \pi \)) 
that are spatiotemporally correlated with the localized amplitude structures 
observed in the corresponding amplitude plots. This indicates the 
formation of coherent, solitary wave structures with non-trivial phase dynamics.
The increasing complexity and number of these phase defects with 
higher \( q \) directly correspond to the higher density 
of wave packets formed. The phase evolution confirms the 
breakdown of the initial plane wave's coherence and the 
establishment of a chaotic, turbulent-like state dominated by nonlinear interactions.

The energy density plots in Fig.~\ref{fig02_ene} provide a 
consolidated view of the system's dynamics, showing where energy becomes localized due to MI.
For \( q = 1.0 \) (Fig.~\ref{fig02_ene}$a_{1}, \, b_{1}$), 
the energy remains largely delocalized and stable over time, 
with the initial perturbation dispersing without forming 
significant localized structures. This confirms the system 
is in a stable propagation regime for this wave number.
With increasing \( q \) (Figs.~\ref{fig02_ene}$a_{2-6}, \, b_{2-6}$), 
the onset of MI is marked by a clear break-up of the uniform energy 
background into well-defined, stable wave packets. The energy becomes 
concentrated in discrete, particle-like excitations that persist over time. 
The number of these energy packets increases with \( q \), 
and their trajectories indicate a complex interplay of propagation and scattering.
The stability and persistence of these high-energy regions 
confirm the formation of robust nonlinear localized modes, 
such as lattice solitons or breathers, which result from a 
balance between the system's dispersion, nonlinearity, and spin-orbit coupling.

The three-dimensional representations in Fig.~\ref{fig02_3d} 
offer a volumetric perspective of the wave amplitude dynamics, 
vividly illustrating the transition from stability to MI. 
The yellow and blue surfaces correspond to $|\phi_n^{+}(t)|$ 
and $|\phi_n^{-}(t)|$, respectively.
For \( q = 1.0 \) (Fig.~\ref{fig02_3d}$a_1, \, b_1$), 
the nearly flat and uniform surfaces of both $|\phi_n^+|$ and $|\phi_n^-|$ 
visually confirm the stable propagation of the 
initial plane wave, with only minor ripples from the small perturbation.
As \( q \) increases within the MI regime (Figs.~\ref{fig02_3d}$a_{2-6}, \, b_{2-6}$),
 the dynamics undergo a dramatic transformation. 
 The flat surfaces break up into a series of well-defined,
  coherent peaks and troughs. These structures represent 
  the formation of stable, localized wave packets. 
  The 3D view clearly shows the soliton-like nature 
  of these excitations, characterized by their persistent 
  amplitude and defined trajectories in the space-time landscape.
The increasing complexity and number of these localized 
excitations with higher \( q \) is directly visible, 
with the system evolving into a dense lattice of interacting nonlinear waves. 
The symmetric development in both components is immediately apparent, 
underscoring the balanced nature of the nonlinear interactions and the role of the SOC in the instability process.}

Figure~\ref{fig03} shows the instability region for $Q$ vs $g_{12}$ 
and in the absence of SO coupling ($\gamma=0$) in the staggered mode ($q=\pi$).
The top and bottom rows are for $g=1$ and $g=2$, respectively. 
Also, the left and right columns are for $\zeta_1$ and $\zeta_2$, respectively. 
In all four panels, the MI regions have symmetry 
with respect to $g_{12}$. 
For $g=2$, the symmetry occurs for both the 
perturbed wave-number $Q$ and the inter-component
interaction $g_{12}$. 
But, for $g=1$, the symmetry happens only concerning the inter-component interaction. 
Further, for $g=1$ (in top panels), the MI region exists up to $Q \leq \pi/2$. But, for $g=2$, it expands up to $Q = \pi$. 
Also, we should notice that the $\zeta_i$ amplitudes are $2$ for $g=1$ and $4$ for $g=2$.

\begin{figure}[t!]
\includegraphics[width=.21\textwidth]{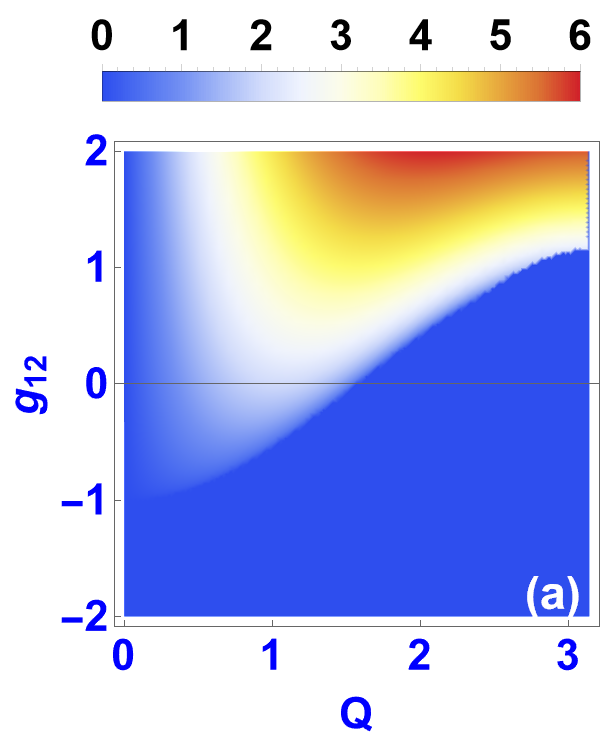} %
\includegraphics[width=.21\textwidth]{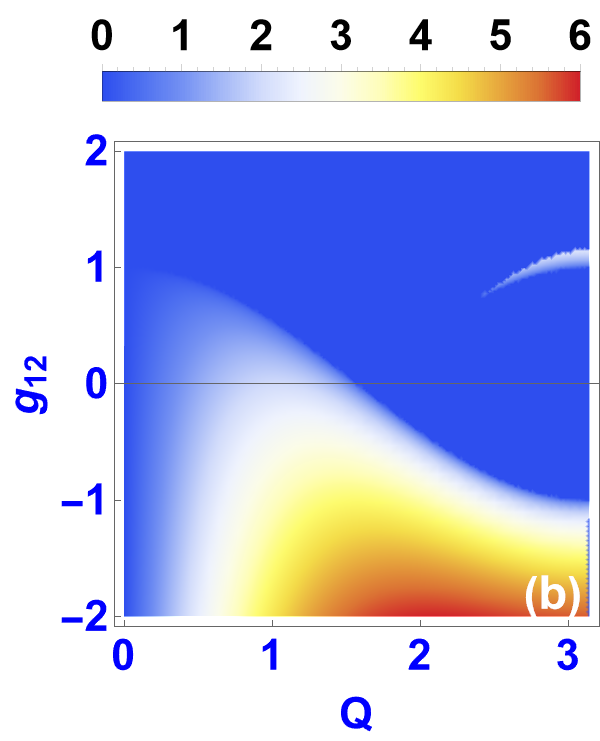}
\par
\includegraphics[width=.21\textwidth]{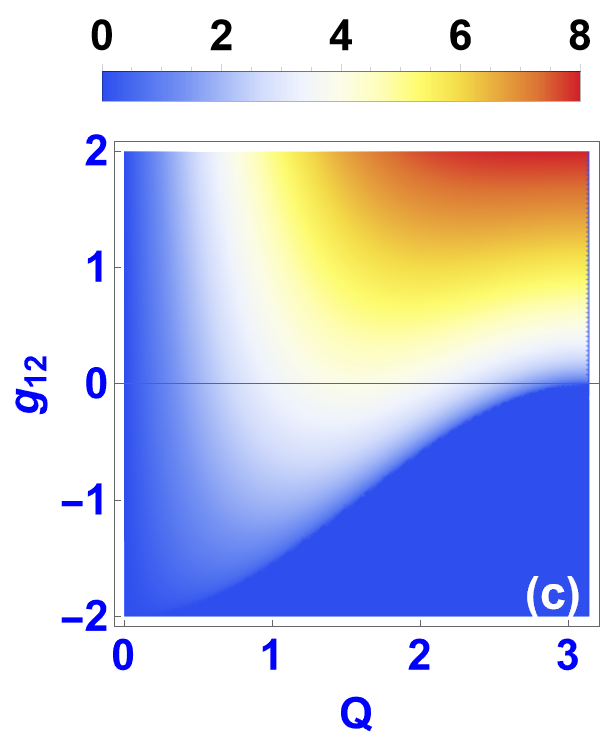} %
\includegraphics[width=.21\textwidth]{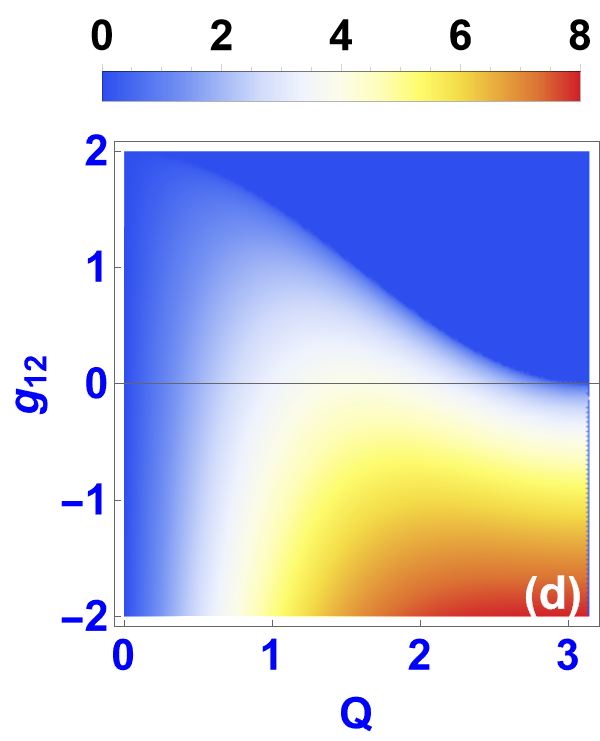}
\caption{(color online) The MI gain for $\gamma=1$ and all the other 
parameters are the same as in Fig.~\ref{fig03}.}
\label{fig04}
\end{figure}
\begin{figure}[t!]
\includegraphics[width=.21\textwidth]{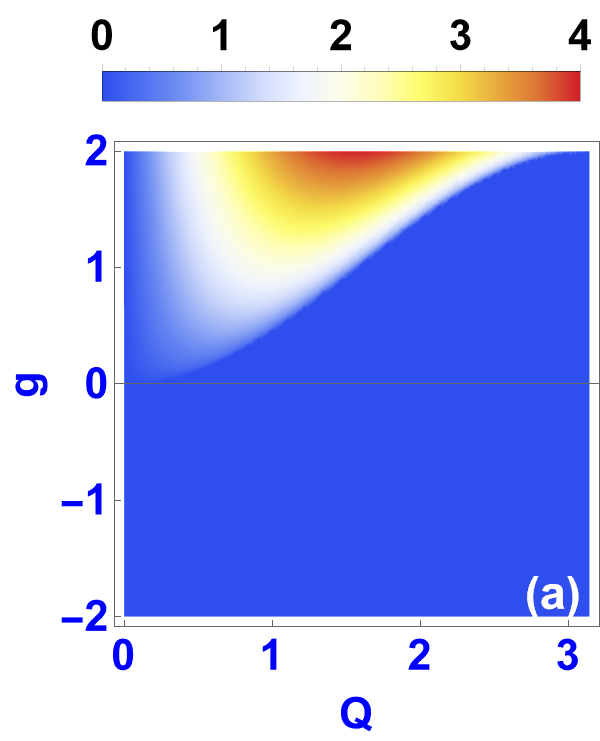} %
\includegraphics[width=.21\textwidth]{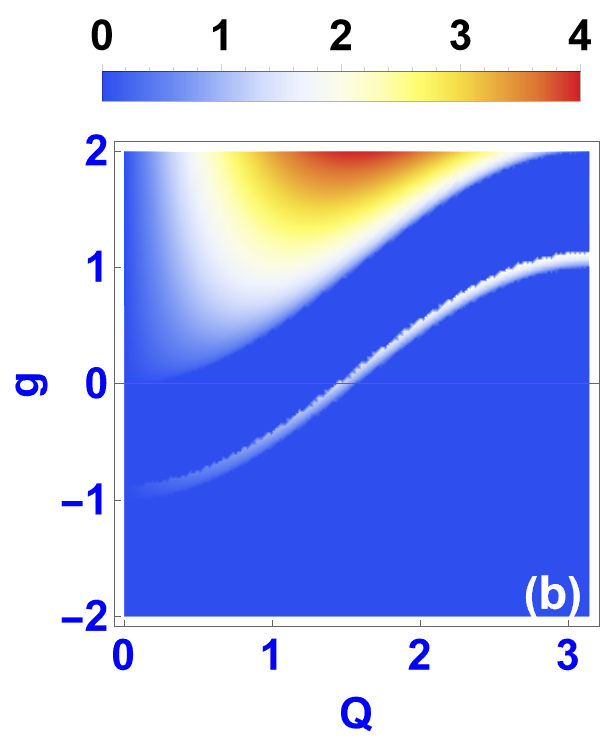}
\par
\includegraphics[width=.21\textwidth]{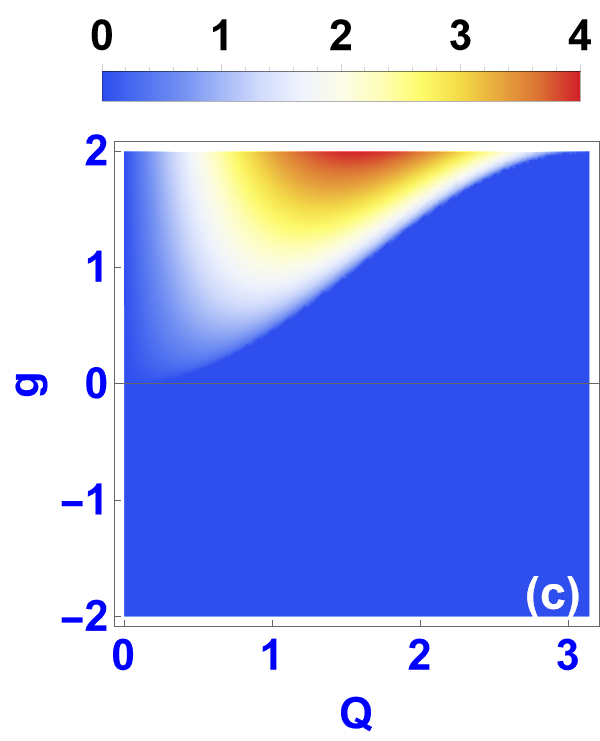} %
\includegraphics[width=.21\textwidth]{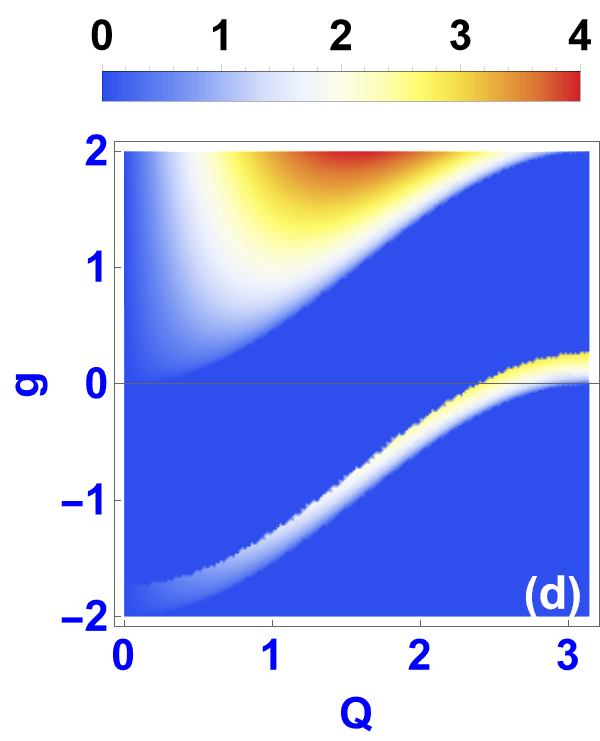}
\caption{(color online) The MI gain for $Q$ vs $g$ and for  
$\gamma=0$ in staggered mode. Left and right panels are for 
$\Omega_1$ and $\Omega_2$, respectively. Top and bottom panels 
are for $g_{12}=1$ and $g_{12}=2$, respectively.}
\label{fig05}
\end{figure}
 
The existence of the MI regions drastically changes when we switch on the 
effect of the SO coupling, as illustrated in Fig.~\ref{fig04}. Here, the 
strength of the SO coupling is given by $\gamma=1$ and all the other 
parameters are the same as in the corresponding panels of Fig.~\ref{fig03}. 
It is clear in Fig.~\ref{fig04} that no symmetry exists in any panels 
like in Fig.~\ref{fig03}. But, the amplitudes are increased three times in 
the case of $g=1$ and two times for $g=2$. Further, the maximum amplitude 
region exists in the repulsive and attractive inter-component interaction 
region for $\zeta_1$ and $\zeta_2$, respectively. Also, if one compares the 
top panels with the bottom panels, the bottom panels have more MI region 
for both $\zeta_1$ and $\zeta_2$. This is because of the difference in the 
strength of the intra-component interaction, $g$ in these cases.

Next, in Fig.~\ref{fig05}, we present and analyze the effects of the 
intra-component interaction over the perturbed wave number $Q$, considering
$\gamma=0$, such that we can verify the effect on the MI of the non-linear 
parameters. In this case, the results are for $\Omega_1$ [panels (a) and (c)]
and $\Omega_2$ [panels (b) and (d)], by assuming $g_{12}=1$ [panels (a) and (b)] 
and 2  [panels (c) and (d)], with $g$ varying from -2 to +2.
In the case of $\Omega_1$, the MI gain occurs only in the repulsive 
intra-component interaction region ($g>0$), with the corresponding region 
decreasing for increasing values of $Q$. Also, the effect of $g_{12}$ going from 1 to 2
is not visible. For $\Omega_2$, we have one branch of results quite similar to the
case of the $\Omega_1$; but another branch of MI (for lower values of $g$) can also 
be verified, being more pronounced for $g_{12}=2$ than for $g_{12}=1$.
Further, there is not much variation either in the amplitudes or in the areas 
of the MI regions.

\begin{figure}[!t]
\includegraphics[width=.21\textwidth]{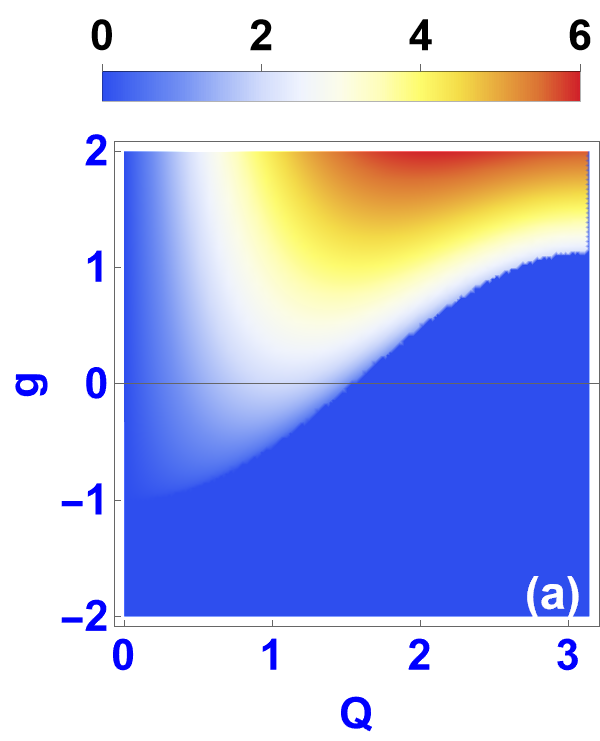} %
\includegraphics[width=.21\textwidth]{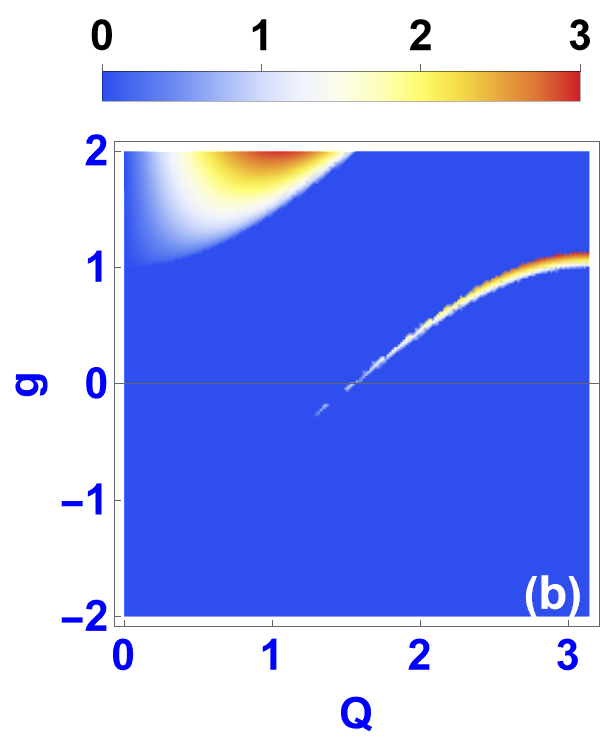}
\par
\includegraphics[width=.21\textwidth]{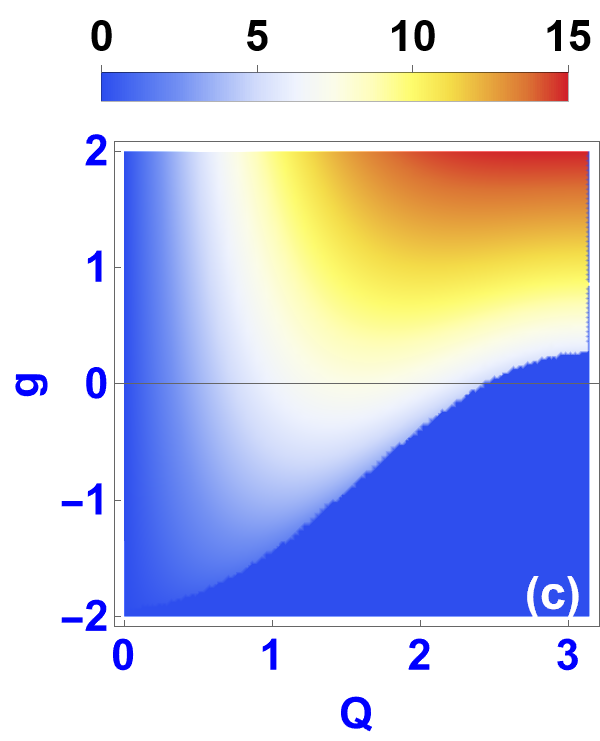} %
\includegraphics[width=.21\textwidth]{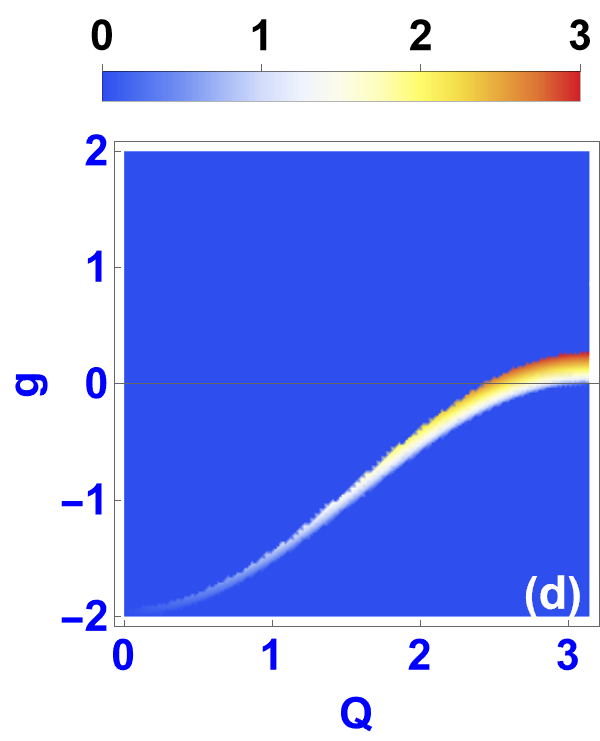}
\caption{(color online) The MI gain for $\gamma=0.5$ and all the other 
parameters are same as in Fig.~\ref{fig05}.}
\label{fig06}
\end{figure}

\begin{figure}[t!]
\includegraphics[width=.21\textwidth]{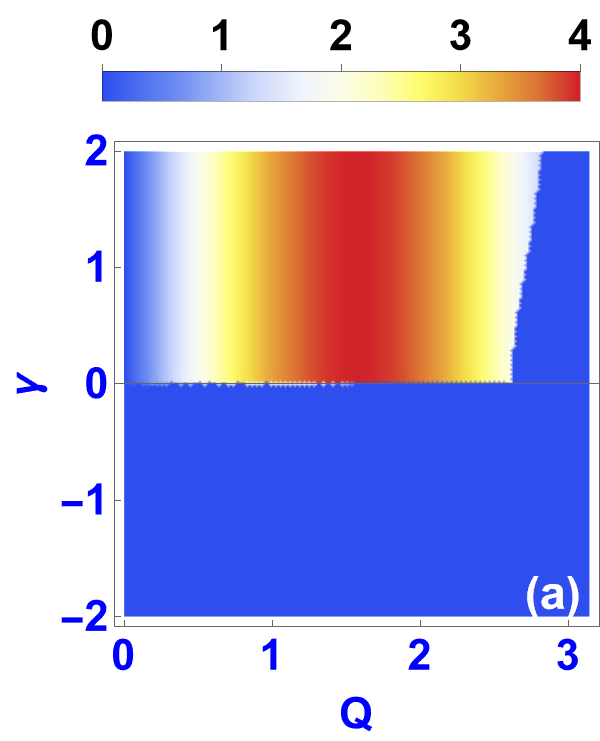} %
\includegraphics[width=.21\textwidth]{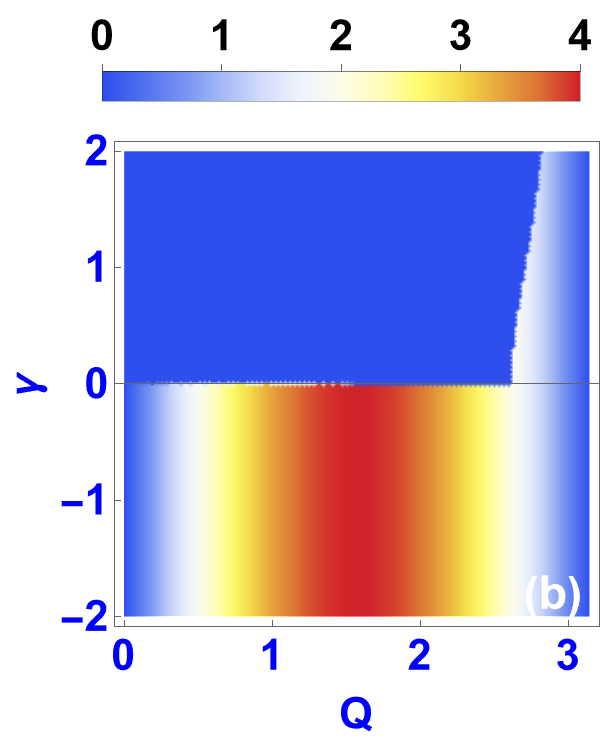}
\par
\includegraphics[width=.21\textwidth]{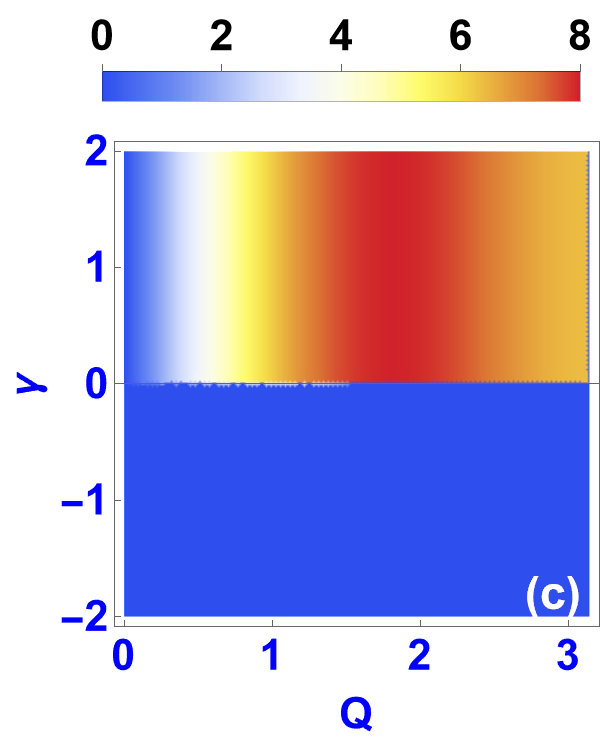} %
\includegraphics[width=.21\textwidth]{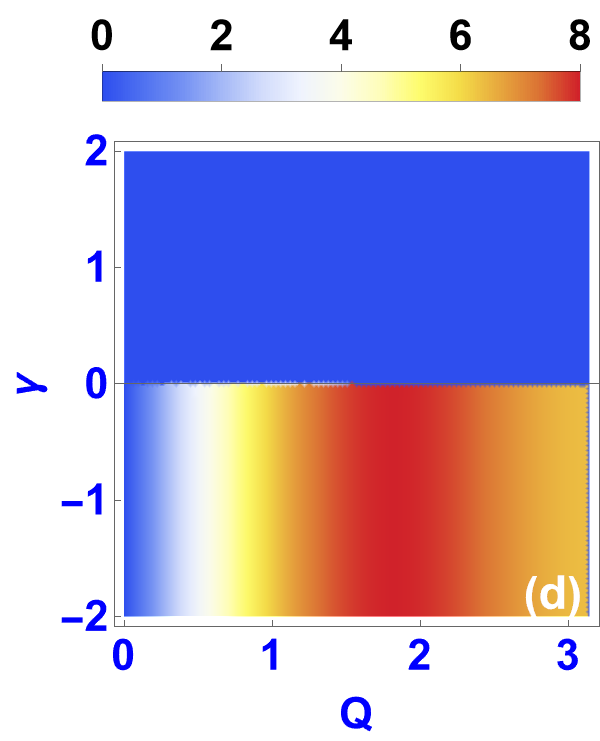}
\caption{(color online) Two-dimensional (2D) plot showing the MI 
gain in staggered mode for $Q$ vs $\gamma$ and for  $g=1$. Left and 
right panels are for $\Omega_1$ and $\Omega_2$, respectively. 
Top and bottom panels are for $g_{12}=1$ and $g_{12}=1.5$, respectively.}
\label{fig07}
\end{figure}
\begin{figure}[t!]
\includegraphics[width=.21\textwidth]{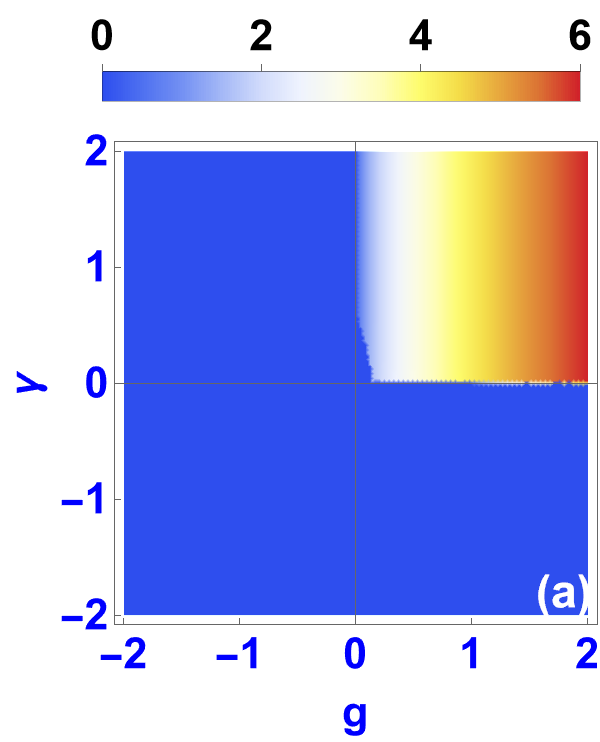} %
\includegraphics[width=.21\textwidth]{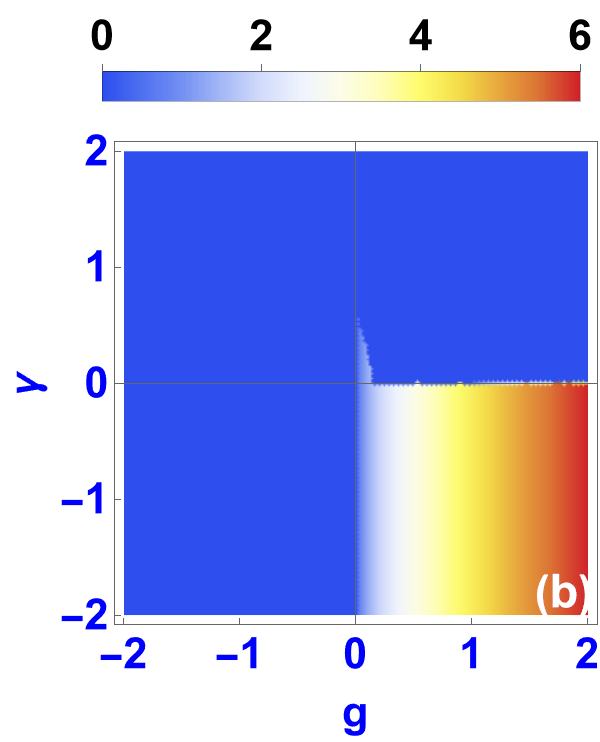}
\par
\includegraphics[width=.21\textwidth]{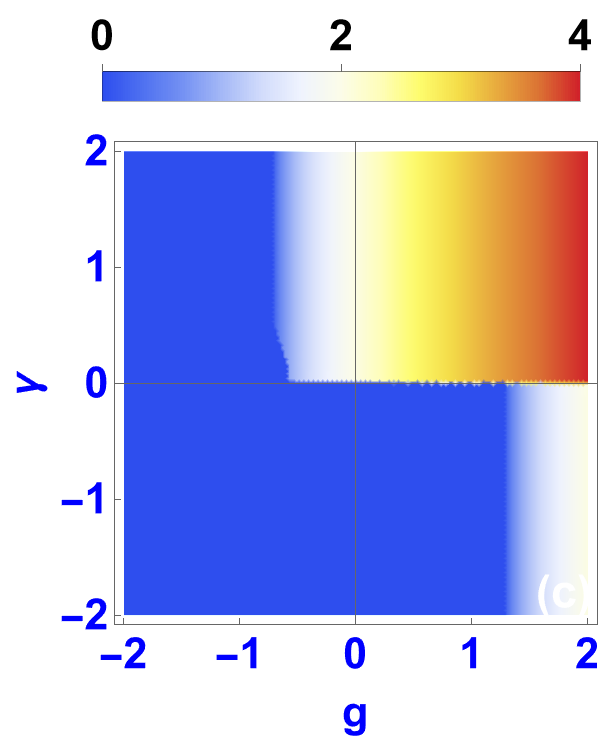} %
\includegraphics[width=.21\textwidth]{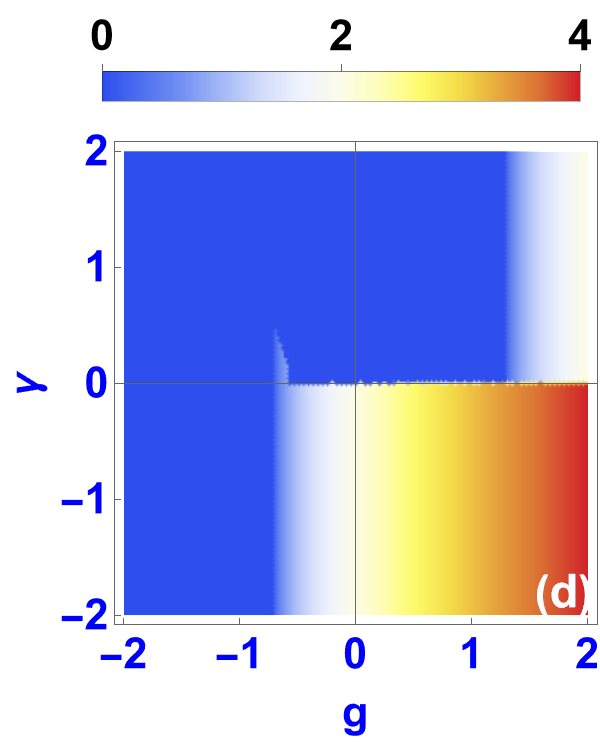}
\caption{(color online) Two-dimensional (2D) plot showing the MI gain in staggered 
mode for $g$ vs $\gamma$ and for $g_{12}=1$. Left and right panels are for $\Omega_1$ 
and $\Omega_2$, respectively. The upper and bottom panels are for 
$Q=\pi/2$ and $Q=\pi/4$, respectively. }
\label{fig08}
\end{figure}
The results given in Fig.~\ref{fig06} show the effect of the SO coupling when 
$\gamma=0.5$ on the MI, when considering the MI gain in parametric regions
given by the intra-species parameter $g$ versus $Q$. 
For comparison with the results given in Fig.~\ref{fig05}, 
all the other parameters are kept the same, keeping the 
correspondence between the panels. Therefore, panels (a) and (b)
are for the inter-species $g_{12}=1$, with the panels (c) and (d) for 
$g_{12}=2$. Panels (a) and (c) are for $\Omega_1$, with panels (b) and (d)
for $\Omega_2$.
As verified by comparing the respective panels of Figs.\ref{fig05} and \ref{fig06}, 
the main effect when switching on the SO coupling parameter
is verified by the MI increasing in case of $\Omega_1$, with a
substantial decreasing for $\Omega_2$.

In Fig.~\ref{fig07}, the SO coupling $\gamma$ is being
varied from negative to positive values, as function of $Q$,
considering $g=1$ (in all the cases), with $g_{12}=1$ [panels (a)
and (b)] and $g_{12}=1.5$ [panels (c) and (d)]. With these 
results, we are verifying that both solutions $\Omega_1$ 
[panels (a) and (c)] and $\Omega_2$ [panels (b) and (d)] provide results for
the MI that are complementary. These results are implying that
the stable regions [when the $\rm{Im}(\Omega_{i=1,2})=0$] are 
quite limited, being close to $Q=0$, in all the cases; and 
close to $Q=\pi$ in case that $g_{12}=1$. In most of the 
cases, when $\rm{Im}(\Omega_{1})=0$ we have 
$\rm{Im}(\Omega_{2})\ne 0$, and vice versa.
With the results given in Fig.~\ref{fig08}, we are investigating
the behavior of the results previously shown in Fig.~\ref{fig07},
at two specific cases with $Q=\pi/2$ and $Q=\pi/4$, by considering
fixed $g_{12}=1$ and varying $g$ from negative to positive values.

\section{Conclusions} \label{sec5}
We have studied the conditions to emerge modulational instabilities 
in a Dresselhaus-Rashba spin-orbit coupled binary system confined 
in deep optical lattices.
Considering two hyperfine states of the same atom, the spin-orbit 
coupling is assumed between inter-species, with Rabi coupling 
occurring among the intra-species. With the coupled system confined 
in a deep optical lattice, the stability analysis is performed 
through the tight-binding model, in which the interactions occur 
only among the closest sites. With the discrete optical lattice  
sites described by plane waves, with the sites affected by the 
wave number $q$ and frequency $\omega$, it was further considered
they are subject to periodic perturbations, having wave number 
$Q$ and frequencies $\Omega$. As a result of our study of this 
coupled system, four general complex solutions $\Omega_{i=1\to4}$
are obtained for the dispersion relations. The corresponding
growth rate instabilities, called gain, 
$\zeta_i=2 |\text{Im} (\Omega_i)|$ are obtained, with the
results being presented for several significant spin-orbit conditions, 
with the main focus on the staggered modes given by $q=\pi$. 
Noticeable was the fact that the dispersion relations came out 
independently on the Rabi constant in this inter-SOC case, being the main
difference in relation to the previously intra-SOC 
investigation. 
In our present analysis and approach, our aim was mainly concerned to 
investigate possible different effects, and corresponding parameter 
dependences which could emerge in comparison with previous studies considering
intra-SOC systems. As known, in principle there is no particular advantages 
in using one experimental setup in relation to the other, such that it will
be dependent on the desired effects that are being verified.
These analytical findings were robustly supported by direct numerical 
simulations of the full discrete system. The evolution of wave amplitudes, 
phase, and energy provided a clear visualization of the instability onset, 
showcasing the breakup of initial plane waves into robust, localized structures. 
The excellent agreement between the linear stability analysis and the nonlinear 
numerical results confirms the validity of our theoretical approach. This work not 
only deepens the understanding of MI in complex, discrete systems with spin-orbit 
coupling but also provides a pathway for controlling instability and generating 
specific nonlinear waveforms in experimental settings through precise 
manipulation of the system's parameters.

\noindent{\small {\bf Acknowledgements:}
SS and LT acknowledges the Funda\c c\~ao de Amparo 
\`a Pesquisa do Estado de S\~ao Paulo [Contracts 2020/02185-1, 2017/05660-0, 2024/04174-8, 2024/01533-7]. 
LT also acknowledges partial support from Conselho Nacional de Desenvolvimento 
Cient\'\i fico e Tecnol\'ogico (Procs. 304469-2019-0)}.


\end{document}